\documentclass[a4paper,aps,pre,floatfix,twocolumn,nofootinbib,showpacs,superscriptaddress]{revtex4-1}
\usepackage[utf8]{inputenc}
\usepackage[T1]{fontenc}

\usepackage[pdftex]{graphicx}% Include figure files
\usepackage{dcolumn}% Align table columns on decimal point
\usepackage{bm}% bold math
\usepackage{amssymb}
\usepackage{pifont}
\usepackage{amsmath}
\usepackage{times}
\usepackage[normalem]{ulem}
\usepackage[english]{babel}
\usepackage{microtype}
\usepackage{epstopdf}
\usepackage{latexsym}
\usepackage{keyval}
\usepackage{ifthen}
\usepackage{moreverb}
\usepackage{color}

\usepackage[hidelinks,unicode=true]{hyperref}
\hypersetup{colorlinks=true,linkcolor=blue,urlcolor=blue,citecolor=blue,pdfhighlight=/N}

\usepackage{lipsum}

\newcommand{\spps}{\mathcal{S}}

\newcommand{\vel}{ \boldsymbol{v} }

\begin{document}

\title{Phase transitions on a class of generalized Vicsek-like models of collective motion}

\author{Pau Clusella}
\affiliation{Departament de F\'{\i}sica, Universitat Polit\`ecnica de
Catalunya, Campus Nord B4, 08034 Barcelona, Spain}

\author{Romualdo Pastor-Satorras}
\affiliation{Departament de F\'{\i}sica, Universitat Polit\`ecnica de
Catalunya, Campus Nord B4, 08034 Barcelona, Spain}
\date{\today}

\begin{abstract}
  Systems composed of interacting self-propelled particles (SPPs) display
  different forms of order-disorder phase transitions relevant to collective
  motion.  In this paper we propose a generalization of the Vicsek model
  characterized by an angular noise term following an arbitrary probability
  density function, which might depend on the state of the system  and thus
  have a multiplicative character.  We show that the well established
  vectorial Vicsek model can be expressed in this general formalism by
  deriving the corresponding angular probability density function, as well
  as we propose two new multiplicative models consisting on a bivariate
  Gaussian and a wrapped Gaussian distributions.  With the proposed
  formalism, the mean-field system can be solved using the mean resultant
  length of the angular stochastic term.  Accordingly, when the SPPs
  interact globally, the character of the phase transition depends on the
  choice of the noise distribution,  being first-order with an hybrid
  scaling for the vectorial and wrapped Gaussian distributions, and second
  order for the bivariate Gaussian distribution.  Numerical simulations
  reveal that this scenario also holds when the interactions among SPPs are
  given by a static complex network.  On the other hand, using spatial
  short-range interactions displays, in all the considered instances, a
  discontinuous transition with a coexistence region, consistent with the
  original formulation of the Vicsek model.
\end{abstract}

\maketitle
%
%\begin{quotation}
%  In 1995 Vicsek \emph{et al.} proposed a simple model to study the
%  emergence of collective motion in a system composed of a large numbers of
%  self-propelled particles (SPPs).  Since then, the study of models
%  displaying flocking phenomena has grown over time.  In the case of SPPs
%  two main elements must be taken into account: the interaction rule
%  according with a SPPs tends to align, and the stochastic source that
%  difficults the perfect alignment.  In this paper we generalize the
%  formulation of the original Vicsek model in order to account for arbitrary
%  distributions for the stochastic term, including multiplicative effects.
%  The simplicity of the proposed formalism provides a unifying framework to
%  study the effects of noise in Vicsek-like models of SPPs, giving a general
%  expression for the mean-field system, as well as it outlines the stricking
%  differences on the nature of the transition depending on how the
%  interactions between particles are established.
%\end{quotation}
%
\section{Introduction}

The self-organized movement of groups of interacting animals can lead to
complex spatio-temporal patterns on widely different time and length scales,
ranging from the migration of mammal herds, the flocking of birds, the
milling and flocking of fish schools, the trailing of marching insects and
the swarming of flying insects, or the migration of cells and
bacteria~\cite{Couzin2003,Sumpter2006,sumpter2010}. The study of this kind
of phenomena has attracted a great deal of attention in the last decades,
leading to the new interdisciplinary field of research denominated
collective motion~\cite{Giardina2008,vicsek2012,Lopez2012}. The field of
collective motion has greatly advanced recently by the gathering of a
growing body of experimental evidence, but its main impulse has been due to
the study of a large variety of models, aimed at either reproduce the
patterns of movement of
animals~\cite{19821081,Reynolds:1987:FHS:37402.37406} or to understand the
ultimate mechanisms driving the collective behavior of animals. From the
biological point of view, such models tend to take into account the detailed
physiological and behavioral properties of the interacting animals. From a
statistical physics point of view, on the other hand, the observation of the
self-organized nature of collective motion in a wide range of scales has
prompted the formulation of minimal models of self-propelled particles,
inspired in the concept of universality~\cite{goldenfelt1992} and based in
simple rules of movement and alignment.

The first and most prominent of such minimal models is the one proposed in
1995 by Vicsek and coworkers~\cite{Vicsek1995PRL,Ginelli2016}. This
so-called \emph{original} Vicsek model is defined in terms of a set of
overdamped self-propelling particles (SPPs) that evolve in discrete time  on
a continuous Euclidean space of two dimensions, and that are fully
characterized by their time-dependent position $\boldsymbol{r}_j(t)$ and
velocity $\boldsymbol{v}_j(t)$, assumed to have a constant modulus $\vert
\boldsymbol{v}_j(t) \vert = v_0$. The SPPs interact among them by attempting
to align the direction of their velocities to the average velocity of a set
of nearest neighbors.  Full alignment is hindered by a source of noise of
strength $\eta$, that represents the difficulties animals may have to
identify the actual state of motion of their nearest neighbors. 

The interest of the original Vicsek model for the physics community lies in
the fact that it predicts the presence of a dynamic order-disorder
(\emph{flocking}) phase transition at a critical value of the noise
intensity, $\eta_c$, separating an ordered phase located at low noise ($\eta
< \eta_c$) in which the SPPs are aligned and travel coherently in a common
average direction, from a disordered phase at high noise $(\eta > \eta_c)$,
in which SPPs move independently of each
other~\cite{Vicsek1995PRL,Ginelli2016}.  Hence, the Vicsek model allows to
relate the collective motion of animals to the well-known features of
standard phase transitions in condensed matter~\cite{yeomans}. This interest
has prompted the study of a large number of variations, intended to include
different aspects of the behavior of real animals, and to explore their
effects on the properties of the ensuing flocking transition. Among those
Vicsek-like models, we can mention variants studying the effects of
difficulties in processing information~\cite{Gregoire2004PRL}, non-metric
neighborhoods~\cite{Ginelli2010PRL}, interactions mediated by social
networks~\cite{Bode2011,Miguel2018PRL}, restrictions in the field of
vision~\cite{PhysRevE.84.046115}, or nematic alignment~\cite{PRE2012Ngo}.

In this paper we intent to provide a common framework for the study and
classification of some of these variants by proposing a general class of
Vicsek-like models that focuses in the role of the stochastic noise that
leads individuals to deviate from the average direction of their neighbors.
Inspired in the original formulation, an individual in our model moves at
constant speed $v_0$,  aligning its direction of motion with the average
velocity of a set of defined neighbors. The alignment is affected by a
source of random angular noise, that we choose to depend on a noise strength
parameter $\eta$ and might also depend on the local polarization in the
vicinity of the individual, measured as the modulus of the average velocity
of its neighbors.  The model is thus completely defined in terms of the set
of nearest neighbors and the distribution of the angular noise, and can
therefore be mapped to different variations of the Vicsek model.  The
analysis of these mappings allows us to focus on the particular case of what
we call models with \emph{multiplicative noise}, that are those in which the
angular noise distribution depends on the ratio of the noise intensity and
the local polarization.  

While the full analysis of the models can be complex depending on the
particular form of the direction distribution, we develop a general
mean-field theory that  allows to obtain preliminar information on the
nature of a possible flocking transition, given by an order parameter
identified by the modulus of the average velocity of the system (the
polarization).  As we observe, at the mean-field level this transition can
be of first or second order, depending on the details of the angular noise
distribution. For first order transitions, the order parameter shows a
characteristic jump and an additional critical singularity in the ordered
phase, which is a signature of a so-called \emph{hybrid phase
transitions}~\cite{Dorogovtsev2006,Goltsev2006}. 

Beyond the mean-field level, we study the behavior of our model with
multiplicative noise in the case of interactions mediated by a static
complex network~\cite{Newman10}, representative of social
interactions~\cite{Bode2011,Miguel2018PRL} and in the case of a
two-dimensional space, with metric interactions~\cite{Vicsek1995PRL}. In
networks, the order of the transition is preserved with respect of the
mean-field prediction, and still shows signatures of a hybrid nature. In the
more realistic case of spatial metric interactions, we observe that in the
thermodynamic limit all prescriptions of multiplicative noise lead to a
discontinuous transition, in agreement with the behavior of the standard
Vicsek model~\cite{chate2019dry}.

\section{A class of generalized Vicsek-like model}
\label{sec:generalizedmodel}
We consider a class of models of flocking dynamics defined in terms of  $N$
self-propelled particles (SPPs), $\spps = \{ 1, 2, \ldots, N \}$, moving in
a 2D medium defined as a square of size $L$ endowed with periodic boundary
conditions. Particles are characterized by a position in space
$\boldsymbol{r}_j(t)=(x_j(t) , y_j(t))$, and a velocity vector, represented
as a complex number, $\vel_j(t)=v_0 \exp[i\theta_j(t)]$, where $v_0$ is  the
particle speed, assumed constant, and  $\theta_j(t)\in[0,2\pi)$ is the
direction of motion. The SPPs undergo an overdamped dynamics, and their
position is updated in a discrete time framework as~\cite{Ginelli2016}
\begin{equation}
        \boldsymbol{r}_j(t+1)=  \boldsymbol{r}_j(t)+v_0 e^{i\theta_j(t+1)}  \;.
\end{equation}
In their movement, the SPPs interact among them by trying to align their
velocity along the average velocity of a set of other particles in their
close neighborhood. The alignment dynamics is implemented by selecting at
time $t$ a set of neighboring particles, $\mathcal{N}_j(t)$, around particle
$j$, and computing their average velocity
\begin{equation}
  \boldsymbol{u}_j(t)=\frac{1}{|\mathcal{N}_j(t)|}\sum_{k\in\mathcal{N}_j(t)}
  e^{i\theta_k(t)} \equiv a_j(t) e^{i\Theta_j(t)}\;, 
  \label{eq:def_local_pol}
\end{equation}
where $|\mathcal{N}_j(t)|$ denotes the number of neighbors in the set
$\mathcal{N}_j(t)$.  In Eq.~\eqref{eq:def_local_pol} we have defined
$a_j(t)$ as the modulus of the neighbor's average velocity, and
$\Theta_j(t)$ as the orientation of this average velocity. The modulus
$a_j(t)$ can be interpreted as an instantaneous  measure of local order
(polarization) in the flock, whose global counterpart is the instantaneous
polarization
\begin{equation}
  \phi(t) = \frac{1}{N} \left| \sum_{j=1}^N e^{i\theta_j(t)} \right|,
  \label{eq:orderparam_temp}
\end{equation}
the time average of which,
\begin{equation}
  \phi = \lim_{T \to \infty} \frac{1}{T} \int_0^T \phi(t) \; dt,
\end{equation}
plays the role of the order parameter in the flocking
transition~\cite{Vicsek1995PRL}.

The orientational dynamics of the SPPs is represented by the dynamical rule
for the direction $\theta_j(t)$, that is given by
\begin{equation}
	\theta_j{(t+1)}=\Theta_j(t) + \xi_j,
  \label{eq:model}
\end{equation}
where $\xi_j$ are uncorrelated random angular
variables~\cite{mardia2009directional} with support in the interval
$[-\pi,\pi)$. That is, the SPPs follow the direction of the average local
polarization, with a perturbation given by the angular noise $\xi$. These
random angles are extracted from a probability density function (PDF),
$P(\xi; a, \eta)$, that, in the most general scenario, we allow to depend on
the local state of order of the system, as given by the local polarization
$a$, as well as on a parameter $\eta$, defined in the range $\eta \in [0,
\eta_\mathrm{max}]$, that controls the width of the PDF and thus gauges the
level of noise applied to the system. The natural conditions to impose to
the distribution $P(\xi; a, \eta)$ are the following: 
\begin{enumerate}
  \item The PDF is symmetric and centered in zero, $P(\xi; a, \eta) =
    P(-\xi; a, \eta)$;

  \item In the limit $\eta \to 0$, we impose that $P(\xi; a, \eta)$ tends to
    a Dirac delta function, $\lim_{\eta \to 0} P(\xi; a, \eta) =
    \delta(\xi)$. In this limit, the velocity of each SPP becomes completely
    aligned to the average velocity of its neighbors, which leads to a
    completely ordered phase, with all SPPs moving in a common direction; 

  \item In the limit $\eta \to \eta_\mathrm{max}$, we expect the systems to
    behave in a completely random way, so we impose $\lim_{\eta \to
    \eta_\mathrm{max}} P(\xi; a, \eta) = 1/(2\pi)$. In this limit, the
    system reaches a disordered phase.
\end{enumerate}
From these conditions, it is expected that the models will experience a
flocking transition between the ordered and the disordered phase at some
transition point $0 < \eta_c < \eta_\mathrm{max}$.

In this general class of Vicsek-like models, the elements are thus defined
in terms of both the election of the neighborhood of interacting particles,
$\mathcal{N}_j(t)$, and the PDF of the angular noise characterizing the
fluctuations of the direction of motion with respect to the local average.
The  usual choice for the set of interacting neighbours is the metric one in
the original Vicsek model~\cite{Vicsek1995PRL}, consisting of a circle
centered at $\boldsymbol{r}_j(t)$ with radius $r_0$, i.e.
\begin{equation}
  \mathcal{N}_j(t)=\{k \in \spps \; | \; \| \boldsymbol{r}_j(t) -
  \boldsymbol{r}_k(t)\|\leq r_0 \}\;.
  \label{eq:metric}
\end{equation}
Variations of this rule can impose, for example, a field of vision, in which
interacting neighbors are those within a circle of radius $r_0$ and whose
position forms a limited angle with the heading of the SPP~\cite{Tian2009},
i.e.
\begin{equation}
  \mathcal{N}_j(t)=\{k \in \spps \; | \; \| \boldsymbol{r}_j(t) -
  \boldsymbol{r}_k(t)\|\leq r_0 \; \land \; \vel_k(t) \cdot \vel_j(t) \leq
\alpha v_0^2 \},
\nonumber
\end{equation}
for a given $\alpha < 1$.

Other non-metric choices consider a fixed number of nearest neighbors,
obtained from a Voronoi tesselation~\cite{Ginelli2010PRL} or the set of
fixed neighbors in a  static architecture of connections given by a complex
network~\cite{Aldana2003,Bode2011,Aldana2007PRL,Miguel2018PRL}.  In this
case, for a complex network defined in terms of a static adjacency matrix
$A=(a_{jj})$, taking values $a_{ij} = 1$ when particles $i$ and $j$ are
connected and $a_{ij} = 0$ otherwise~\cite{Newman10}, we have 
\begin{equation}
	\mathcal{N}_j=\{k \in \spps \; |\;  a_{jk}=1 \}\;,
\end{equation}
which is time independent.  Notice that for a fixed pattern of interactions, the
position of each particle is irrelevant for the evolution of the system, and
it is usually not considered~\cite{Bode2011,Miguel2018PRL}.

The second element defining the properties of the model in our framework is
the angular noise distribution $P(\xi; a, \eta)$. In order to gain an
intuition about its possible form, we consider two classic models of
flocking dynamics and present their mapping to our generalized model.

\subsection{Scalar Vicsek model}
\label{sec:scalar}

In the original Vicsek model~\cite{Vicsek1995PRL}, the angular noise is
implemented as a uniform random number in the interval  $[-\eta\pi,
\eta\pi]$, with  $\eta\in[0,1]$.  We thus have in this case
\begin{equation}\label{eq:proboriginal}
  P(\xi;a,\eta) \equiv P(\xi;\eta)=\frac{1}{2\pi\eta} H(\eta
  \pi-\xi) H(\xi + \eta \pi),
\end{equation}
independent of the local polarization, and  where $H(x)$ is the Heaviside
step function.  Since in this case there is no dependence on the local
polarization, we can say that the noise has an \emph{additive} nature, i.e.,
it is independent of the local state of the system. The Vicsek model with
this angular distribution is usually referred to as the Vicsek model with
\emph{scalar} or \emph{intrinsic} noise~\cite{Pimentel2008PRE}.

The scalar Vicsek model represents the case with the simplest distribution
of angles, given by a bounded uniform distribution. More complex and
physically motivated distributions appear when we consider variations of the
original Vicsek model, as we will see next.

\subsection{Vectorial Vicsek model}
\label{sec:vectorial}

Gr\'egoire and H. Chat\'e~\cite{Gregoire2004PRL} introduced a modification
of the original Vicsek model aimed at capturing the errors committed by a
single individual in determining the direction of its neighboring particles.
In this case, velocity directions are updated according to the rule
\begin{equation}\label{eq:vectorial}
  \theta_j{(t+1)}=\arg\left\{
  \frac{1}{|\mathcal{N}_j(t)|}\sum_{k\in\mathcal{N}_j(t)} e^{i \theta_k(t)}
+\eta e^{\chi_j(t)}\right\}
\end{equation}
where  $\chi_j$ are independent uniformly distributed random angles in the
interval $\chi_j\in[0,2\pi)$ and the noise parameter belongs to the range
$\eta \in [0, \infty[$.  This type of noise is usually termed as
\emph{vectorial} or \emph{intrinsic}~\cite{Pimentel2008PRE}. 

In order to map this model to our formalism  we note that, since the random
variables $\chi_j$ are uniformly distributed, they are invariant under
rotations~\cite{mardia2009directional}. Therefore, Eq.~(\ref{eq:vectorial})
is equivalent to
\begin{eqnarray}
  \theta_j{(t+1)} 
  &=& \arg\left\{ a_j e^{\Theta_j(t)}  + \eta
  e^{i(\Theta_j(t)+\chi_j)}\right\} \\ 
  &=& \arg\left\{ e^{i \Theta_j(t)}\left( a_j  +\eta e^{i\chi_j}\right)\right\} \\ 
  &=& \Theta_j(t)+\arg\{a_j+\eta e^{i\chi_j}\}\;.
\end{eqnarray}
Hence, we only need to determine the distribution of $\xi=\arg\{a+\eta
e^{i\chi}\}$ where $a$ and $\eta$ are given parameters (the subindices $j$
are dropped for the sake of simplicity).

\begin{figure}[t]
  \includegraphics[width=0.8\columnwidth]{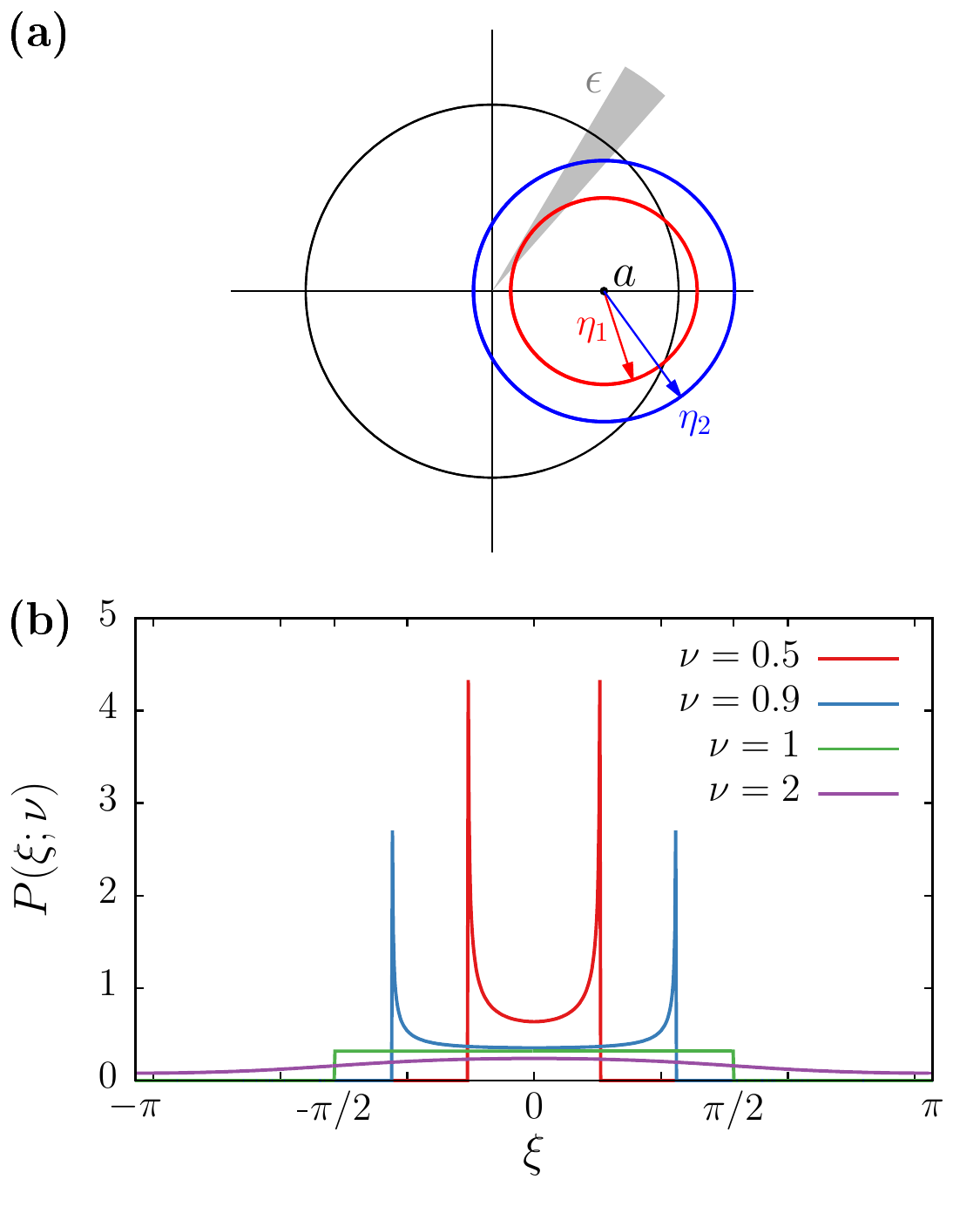}
  \caption{(a) Geometric interpretation of the noise effects in the
    vectorial Vicsek model. The red and blue circles depict the circle $r
    e^{i\xi}=a+\eta e^{i\chi}$ corresponding to two $\eta_1=0.5$ (red) and
    $\eta_2=0.7$ (blue), both cases with a local polarization of $a=0.6$.
    The grey shaded area indicates an interval $\epsilon$
    over which the arclength needs to be computed to
  obtain $P(\xi)$.  (b) Shape of $P(\xi;\nu)$ for different values of
$\nu=\eta/a$, as obtained by Eq.~(\ref{eq:vectorialnu}).}
  \label{fig:vectorial}
\end{figure}

The expression $re^{i\xi}=a+\eta e ^{i\chi}$ defines a circle parametrized by
$\chi$ with center $a$ and radius $\eta$ in the complex plane, see
Fig.~\ref{fig:vectorial}(a).  Separating real and imaginary parts one gets
the implicit equation that defines the circle of possible directions, given
by
\begin{equation}
        r^2=\eta^2+2ra\cos(\xi)-a^2\;,
\end{equation}
which, solving for $r$, leads to
\begin{equation}\label{eq:radius}
        r(\xi)=a\cos(\xi)\pm\sqrt{\eta^2-a^2\sin^2\xi}
\end{equation}
where $\xi\in[-\arcsin(\eta/a),\arcsin(\eta/a)]$ if $\eta<a$ and
$\xi\in[-\pi,\pi]$ otherwise.  Since all the points lying on the circle are
equiprobable, we consider an infinitesimal angular interval.
$[\xi,\xi+\epsilon)$ and compute the arc length over such small interval as
\begin{eqnarray}
  P(\xi; a, \eta)&=&\lim_{\epsilon \to 0}\frac{1}{2\pi\eta
  \epsilon}\int_\xi^{\xi+\epsilon} d\xi \sqrt{r'(\xi)^2+r(\xi)^2}\nonumber\\
                 &=&\frac{\sqrt{r'(\xi)^2+r(\xi)^2}}{2\pi\eta}
\end{eqnarray}
where the normalization constant $\frac{1}{2\pi\eta}$ corresponds to the
total arc length of the circle, and $r'(\xi)$ denotes the first derivative
of $r$ with respect to $\xi$.  Using the expression for $r(\xi)$ in
Eq.~(\ref{eq:radius}) we finally obtain the distribution for the angular
random variables $\xi$
\begin{widetext}
\begin{equation}\label{eq:distribution}
	P(\xi;a,\eta)=
	\begin{cases}
    \displaystyle{\frac{1}{2\pi}\left( 1 + \frac{a \cos(\xi)}{\sqrt{\eta ^2
    - a^2\sin ^2(\xi)}} \right)  } & \mathrm{for} \; \eta>a,\; \xi\in (-\pi,\pi] \\[35pt]
    \displaystyle{\frac{a \cos(\xi)}{\pi \sqrt{\eta^2-a^2\sin^2(\xi)}}}, & 
    \mathrm{for}\; \eta\leq a,\; \xi \in
    {[-\arcsin\left(\frac{\eta}{a}\right),\arcsin\left(\frac{\eta}{a}\right)]}\;.
	\end{cases}
\end{equation}
In this distribution, which does not coincide, to the best of our knowledge,
with any typical angular probability
distribution~\cite{mardia2009directional}, the \emph{multiplicative}
structure of the noise becomes explicit.  In fact, the distribution depends
on the ratio between the local polarization $a$ and the control parameter
$\eta$. Thus, introducing the variable $\nu=\eta/a\geq0$,
Eq.~(\ref{eq:distribution}) can be expressed as
\begin{equation}\label{eq:vectorialnu}
  P(\xi;\nu)=
  \begin{cases}
    \displaystyle{\frac{1}{2\pi}\left( 1 + \frac{ \cos(\xi)}{\sqrt{\nu ^2 -
    \sin ^2(\xi)}} \right)  }, & \nu>1,\; \xi\in (-\pi,\pi] \\[35pt]
    \displaystyle{\frac{\cos(\xi)}{\pi \sqrt{\nu^2-\sin^2(\xi)}}}, &
    \nu\leq1,\; \xi \in {[-\arcsin\left(\nu\right),\arcsin\left(\nu
    \right)]}\;.
        \end{cases}
\end{equation}
\end{widetext}

Fig.~\ref{fig:vectorial}(b) shows the shape of $P(\xi;\nu)$ for various
values of $\nu$.  The piecewise character of $P$ illustrates the strong
effects of the vectorial noise.  If the local polarization of the particle
$a$ is smaller than the noise strength $\eta$ (i.e., $\nu>1$), then the
distribution is unimodal, with $\xi$ ranging within $(-\pi,\pi)$, and tends
to the uniform distribution $P(\xi) = 1/2 \pi$ in the limit $\nu \to
\infty$.  On the other hand, if the local polarization is larger than the
noise strength (i.e., $\nu<1$), then the probability density is concentrated
at the extreme values of the distribution
$\pm\arcsin\left(\frac{\eta}{a}\right)$. In the limit of $\nu \to 0$, the
extremes of the distribution coalesce to yield a delta function, as can be
checked in Fourier space.

\section{Multiplicative models for the angular noise  distribution}
\label{sec:modelsofnoise}

The form of multiplicative noise distribution shown by the Vicsek model with
angular distribution of amplitude (variance) depending on $\nu = \eta / a$,
represents a physically motivated choice, in which the fluctuations of the
particle orientation around the direction of the local polarization,
$\Theta_i(t)$, are an increasing function of the noise strength $\eta$ and a
decreasing function of the local polarization $a_i$, meaning that an
individual has a stronger tendency to follow the average direction of a
group if this group is very coherent (an explicit mathematical relation is
derived in Section~\ref{sec:meanfield}).  For this reason, in the following
we will consider  two additional models characterized by an angular
distribution with a multiplicative nature:  a simple wrapped Gaussian noise,
and a more physically motivated bivariate Gaussian distribution.

\subsection{Wrapped Gaussian noise}

Gaussian distributions represent a natural choice for typical distributions
of noise in mathematical models. In the present case, the noise term has a
distribution with support in the unit circle.  Therefore, the natural
extension of Gaussian noise is provided by a Gaussian distribution defined
at the scalar level and then projected on the unit circle. We thus consider
a model in with the angular noise distribution $P(\xi;a,\eta)$ is a normal
distribution with standard deviation $\sigma=\eta/a$ wrapped on the circle.
The resulting probability density reads~\cite{mardia2009directional},
\begin{equation}
  P(\xi;a,\eta)=\frac{a}{\eta \sqrt{2\pi} }\sum_{m=-\infty}^\infty
  \exp\left[\frac{-a^2(\xi+2\pi m)^2}{2\eta^2}\right]\;.
  \label{eq:wrapped_noise}
\end{equation}
Practically,  one just simulates a usual Gaussian variable $\zeta$
and then wraps it to $[0,2\pi)$ as $\xi=\zeta\;\;(\text{mod } 2\pi)$.

\subsection{Bivariate Gaussian noise}

The inclusion of vectorial noise in the Vicsek model is meant to account, in
average, for the inaccuracy of the single individuals in determining the
exact direction of their neighbors~\cite{Gregoire2004PRL,Chate2008PRE}.
Here we propose a different noise term also aimed to capture these  errors.
The main idea consists of adding a bivariate Gaussian random variable for
each particle in the computation of the mean direction, that can be
interpreted as making a small error in the determination of the orientation
of each neighbor.  Formally, this model can be formulated as
\begin{equation}
  \theta_j(t+1)=\arg\left\{
  \frac{1}{|\mathcal{N}_j|}\sum_{k\in\mathcal{N}_j}\left( e^{i\theta_k(t)} +
\boldsymbol{\zeta}_j^{(k)}(t)\right) \right\},
\end{equation}
where the real and imaginary parts of the complex numbers
$\boldsymbol{\zeta}^{(k)}_j=\alpha^{(k)}_j+i \beta_j^{(k)}$ are independent
Gaussian variables with zero mean and variance $\sigma^2$.   This equals to
consider that the vectors $(\alpha_j^{(k)},\beta_j^{(k)})$ are bivariate
Gaussian variables with covariance matrix $\sigma^2\boldsymbol{I}$.   Then
one can sum the contributions of the different Gaussian terms so that the
system reads
\begin{equation}
  \theta_j(t+1)=\arg\left\{ a_je^{i\Theta_j} +
  \frac{1}{\sqrt{|\mathcal{N}_j|}} \boldsymbol{\zeta}_j(t) \right\}\;, 
\end{equation}
where $\boldsymbol{\zeta}_j=\alpha_j+i\beta_j$ and $(\alpha_j,\beta_j)$ are
also bivariate Gaussian variables with covariance $\sigma^2\boldsymbol{I}$.
The argument of a bivariate Gaussian random variable centered at
$(a_j\cos(\Theta_j),a_j\sin(\Theta_j))$ and standard deviation $\sigma$
follows a projected normal distribution~\cite{mardia2009directional}.  Since
the projected normal distribution is closed under rotations, the generalized
Vicsek model with bivariate Gaussian noise can be written in the form 
\begin{equation}
  \theta_j(t+1)=\arg\left\{
	  a_je^{i\Theta_j}
  %\frac{1}{|\mathcal{N}_j|}\sum_{k\in\mathcal{N}_j} e^{i\theta_k(t)}
\right\}+\xi_j(t),
\end{equation}
where $\xi_j$ is a projected normal angular variable centered at $(a_j,0)$
and covariance matrix $\sigma^2 \boldsymbol{I}/\sqrt{|\mathcal{N}_j|}$.
Notice that the mean direction $\Theta_j$ does not have an impact on the
noise distribution, whereas the local polarization $a_j$ does.

In this particular case there are two sources of multiplicative effects: The
local polarization $a_j$ and the number of neighboring particles
$|\mathcal{N}_j|$.  The study of the impact of the number of neighboring
particles on the system is beyond the scope of this paper and we leave it
for future studies. Instead, we consider a simplified version where the
standard deviation of the variables $\alpha_j$ and $\beta_j$ is just the
control parameter $\eta$.  Accordingly, the probability density for the
random variables $\xi$ as formulated in Eq.~(\ref{eq:model}) is a projected
normal distribution corresponding to a bivariate Gaussian centered at $a$
and with standard deviation $\eta$, which leads to (see Eq.(3.5.48) in
Ref.~\cite{mardia2009directional})
\begin{equation}
  P(\xi;a,\eta)=c_0+\frac{a}{\eta}\cos(\xi)\Phi\left(\frac{a}{\eta}
  \cos(\xi)\right)f\left(\frac{a}{\eta}\sin(\xi)\right)
  \label{eq:bivariate_noise}
\end{equation}
where $c_0=\exp{\left(\frac{-a^2}{2\eta^2}\right)}/(2\pi)$ is a constant and
$\Phi(z)$ and $f(z)$ are the cumulative distribution and probability density
functions of a normally distributed random variable with zero mean and unit
variance, respectively.  It is worth noticing that, again, the distribution
only depends on the parameter, $\eta/a$, which arises naturally by
implementing the projection of the stochastic vectors on the unit circle.

\section{Mean-field theory}
\label{sec:meanfield}

We can gain a preliminar understanding of the properties of our class of
models by means of a mean-field analysis, by which we mean the limit case in
which each SPP interacts with all other particles or, in other words, when
interactions are mediated by a fully connected network. 
In this case, we have 
\begin{equation}
  \Phi=\phi e^{i\alpha}=\frac{1}{N}\sum_{j=1}^N e^{i\theta_j}
\end{equation}
where $\phi=|\Phi|$ denotes the order parameter of the system.  The velocity
of each variable is updated according to
\begin{equation}
	\theta_j(t+1)=\alpha(t)+\xi_j(t)\;,
\end{equation}
thus, one can write an equation for the evolution of the order parameter as
\begin{equation}
	\phi(t+1)=\frac{1}{N}\left|\sum_{j=1}^N e^{i\alpha(t)+i\xi_j(t)}\right|
  =\frac{1}{N}\left|\sum_{j=1}^N e^{i\xi_j(t)}\right|\;.
\end{equation}
The distribution of each $\xi_j$ depends, in principle, on $\eta$ and the
local polarization $a_j$. In the mean-field level, however, $a_j=\phi$.  The
subindices on the previous equation can thus be dropped to obtain
\begin{equation}
  \phi(t+1)=\frac{1}{N}\left|\sum_{j=1}^N e^{i\xi_j(t)}\right| \;.
\end{equation}
Invoking the thermodynamic limit ($N\to\infty$) 
one finally obtains the evolution equation of the mean-field as
\begin{eqnarray}  \phi(t+1) &=& E[\cos(\xi)] =\int_{-\pi}^\pi
  P(\xi;\phi(t),\eta)\cos(\xi) d\xi \\
            &\equiv& \rho(\phi(t),\eta)\;,
\end{eqnarray}
where we are using the fact that the distribution $P$ is symmetric and
centered at zero.  The fixed points of the previous equation correspond to
the stationary states of the mean-field solution, which are obtained as the
solution of the self-consistent mean-field equation
\begin{equation}
  \phi =\rho(\phi,\eta) = \int_{-\pi}^\pi P(\xi;\phi,\eta)\cos(\xi)
  d\xi.
 \label{eq:meanfield}
\end{equation}

In circular statistics the quantity $\rho=E[\cos(\xi)]$ is known as
\emph{mean resultant length} and it is a well studied measure of
concentration~\cite{mardia2009directional}.  Moreover, the \emph{circular
variance} of an angular distribution is usually defined as $V=1-\rho$.  The
fact that, in a globally coupled system, the order parameter of the system
corresponds exactly to the mean resultant length of the noise distribution
$P(\xi)$ is an important result that simplifies the analysis of the
mean-field dynamics.  

In the following, we will consider the mean-field solution of the models
defined by the angular distributions considered above. 

\subsection{Scalar Vicsek model}

The original Vicsek model, characterized by the probability distribution
presented in Eq.~(\ref{eq:proboriginal}), presents the mean-field solution
\begin{equation}\label{eq:rhoA}
  \phi(\eta)=\rho(\eta)=\frac{\sin(\pi\eta)}{\pi\eta}\;,
\end{equation}
which does not undergo any phase transition except at the limiting value
$\eta=1$, for which the system becomes disordered~\cite{Miguel2018PRL}. In
Fig.~\ref{fig:meanfield}(a) we present a plot of the shape of the mean-field
order parameter as a function of $\eta$.

\subsection{Vectorial Vicsek model}

The vectorial Vicsek model, defined by the angular noise distribution
Eq.~\eqref{eq:vectorialnu}, has a mean resultant length for $\nu>1$,
of the form
\begin{equation}
  \rho = \int_{-\pi}^{\pi} d\xi
  \frac{\cos^2\xi}{2\pi\sqrt{\nu^2-\sin^2\xi}} = \frac{1}{2\nu}
  {_2F_1}\left(\frac{1}{2},\frac{1}{2};2,\frac{1}{\nu^2}\right),
\end{equation}
where ${_2F_1}$ is the hypergeometric
function~\cite{abramowitz1965handbook}.  For the
case $\nu\leq1$, denoting $\xi_0 = \arcsin(\nu)$, we obtain 
\begin{equation}
	\rho=\int_{-\xi_0}^{\xi_0} d\xi \frac{\cos^2\xi}{\pi\sqrt{\nu^2-\sin^2\xi}}
	={_2F_1}\left(\frac{1}{2},\frac{-1}{2};1,\nu^2\right)\;.
\end{equation}
Using the fact that $\nu=\eta/\phi$ we obtain the mean-field equation for
the Vicsek model with vectorial noise,
\begin{equation}\label{eq:rhoB}
	\phi = \rho(\phi,\eta)=
	\begin{cases}
		\displaystyle{\frac{\phi}{2\eta}
    {_2F_1}\left(\frac{1}{2},\frac{1}{2};2,\frac{\phi^2}{\eta^2}\right)} & 
    \phi<\eta\\[10pt]
		\displaystyle{{_2F_1}\left(\frac{1}{2},\frac{-1}{2};1,\frac{\eta^2}{\phi^2}\right)}
    & \phi\geq\eta
	\end{cases}
\end{equation}
We notice that this result
was already reported in \cite{Pimentel2008PRE}, obtained using a completely
different approach.  An explicit solution for $\phi=\rho(\phi,\eta)$
is, as far as we know, out of reach. However, in the present case of
multiplicative noise, since the angular noise distribution depends on the
ratio $\nu = \eta/\phi$, we can write the mean-field equation as
\begin{equation}
  \eta = \frac{\eta}{\phi} \rho\left( \frac{\eta}{\phi}\right) \equiv \nu
  \rho(\nu).
  \label{eq:meanfield_numeric_multiplicative}
\end{equation}
As a result, the mean-field solution can be obtained numerically simply by
plotting the noise parameter $\eta(\nu) = \nu \rho(\nu)$, as a function of
$\nu$, and obtaining the order parameter from the expression $\phi =
\eta(\nu) / \nu$.  The red curve in Fig.~\ref{fig:meanfield}(b) shows the
monotonically increasing relation between the circular variance, 
$V=1-\rho$, and the distribution parameter $\nu$, whereas the inset shows
the form of $\eta(\nu)$ as a function of $\nu$.  In
Fig.~\ref{fig:meanfield}(a) we plot the corresponding mean-field solution.
The continuous line indicates a stable fixed point, whereas the dashed line
indicates unstable solutions.  The curves are accompanied by results of
numerical simulations of the mean-field system for $N=10^4$ starting from an
ordered state.

While the original Vicsek model does not present a phase transition in the
mean-field limit,  the vectorial Vicsek model does display one: for large
$\eta$ only the disordered solution exists (i.e., $\phi=0$).  As noise is
reduced, two new solutions appear through a saddle-node bifurcation at
$\eta_c\simeq 0.6715$ (determined numerically), thus corresponding to a
discontinuous phase transition.  Only one of the two new solutions is
stable.  For a range of intermediate values of $\eta$ there is
multistability between the ordered and disordered states.  The value
$\eta_{c'}$ for which the solution $\phi=0$ collides with the unstable state
and loses stability can be obtained analytically as
\begin{equation}
  \frac{d}{d\phi}
  \rho(\phi,\eta_{c'})\biggr|_{\phi=0}=\frac{1}{2\eta_{c'}}{_2F_1}\left(\frac{1}{2},\frac{1}{2};2,0\right)=\frac{1}{2\eta_{c'}}
  = 1\;. 
\end{equation}
Hence for  $\eta<\eta_{c'}=1/2$ only the ordered state is stable.

\subsection{Wrapped Gaussian noise}

For the wrapped Gaussian noise, defined by the angular noise distribution in
Eq.~\eqref{eq:wrapped_noise}, the mean resultant length takes the form (see
Eq.(3.5.63) from Ref.~\cite{mardia2009directional})
\begin{equation}\label{eq:rhoD}
	\rho=\exp \left(-\frac{\eta^2}{2\phi^2}\right),
\end{equation}
which, from the mean-field equation $\phi = \rho(\phi, \eta)$, leads to the
implicit mean-field solution
\begin{equation}\label{eq:etaWG}
	\eta=\phi\sqrt{-2\log(\phi)}\;,
\end{equation}
which is valid only for $\phi > 0$.  However, since, for finite $\eta>0$,
\begin{equation}
	\lim_{\phi \to 0}\rho(\phi,\eta)=0\;,
\end{equation}
we can see that the solution $\phi=0$ exists as a limiting case.

The wrapped Gaussian distribution exhibits a discontinuous phase transition
at the mean-field level. The equation Eq.~\eqref{eq:rhoD} has two branches
of solutions, a stable and an unstable one. Both branches coexist with the
disordered state $\phi=0$, which is stable for all values of $\eta$.  The
ordered state vanishes at the local maximum of $\eta$ in
Eq.~(\ref{eq:etaWG}), which yields a transition point $\eta_c=e^{-1/2}$ with
the jump of the discontinuous transition being also $\phi_c=e^{-1/2}$. In
Fig.~\ref{fig:meanfield}(a) we present a plot of the resulting mean-field
solution for the order parameter $\phi$.  {Also in this case, the (circular)
  variance of the distribution increases with $\nu$, as illustrated by the
  green line in Fig.~\ref{fig:meanfield}(b).

\subsection{Bivariate Gaussian distribution}

The bivariate Gaussian distribution, characterized by the angular noise
distribution Eq.~\eqref{eq:bivariate_noise}, has an associated mean
resultant length~\cite{Kendall1974},
\begin{equation}\label{eq:rhoC}
	\rho =\frac{\phi}{2\eta}\sqrt{\frac{\pi}{2}}e^{-\frac{\phi^2}{4\eta^2}}
	\left[I_0\left(\frac{\phi^2}{4\eta^2}\right) + I_1\left(\frac{\phi^2}{4\eta^2}\right)\right]\;
\end{equation}
where $I_0(z)$ and $I_1(z)$ are modified Bessel functions of the first
kind~\cite{abramowitz1965handbook}.

In this case, the mean-field equation $\phi = \rho(\phi, \eta)$ leads to a
continuous phase transition, as a numerical analysis based on
Eq.~\eqref{eq:meanfield_numeric_multiplicative} shows.  
As indicated by the  blue squares in Fig~\ref{fig:meanfield}(a), upon reducing $\eta$
the disordered state ($\phi=0$) loses stability through a pitchfork
bifurcation, giving raise only to one new meaningful stable state. Although
the mean-field equation is hard to analyze, it is possible to compute the
critical point value $\eta_c$, since it corresponds to the loss of stability
of the disordered solution $\phi=0$. Hence, $\eta_c$ is given by
\begin{equation}
	\frac{d}{d\phi} \rho(\phi,\eta)\biggr|_{\phi=0}=\frac{1}{2\eta_c}\sqrt{\frac{\pi}{2}} = 1
\end{equation}
thus $\eta_c=\frac{1}{2}\sqrt{\frac{\pi}{2}}\simeq 0.627$.
Using this value  it is possible to compute the critical exponent $\beta$ for the phase transition.
Performing a Taylor expansion on both sides of the mean-field equation $\phi=\rho(\phi,\eta)$ for $\phi\ll1$ up to third order
one obtains the relation
\begin{equation}
	1=\frac{\eta_c}{\eta}\left(1-\frac{\phi^2}{8\eta^2}\right)\;.
\end{equation}
Solving for $\phi$ and taking $\eta\to\eta_c$ finally leads to
\begin{equation}
	\phi=\frac{2\sqrt{2}\eta}{\sqrt{\eta_c}}\left(\eta_c-\eta\right)^{\frac{1}{2}}\propto (\eta_c-\eta)^{\frac{1}{2}}\;,
\end{equation}
thus near the transition the order parameter scales with an exponent of
$\beta=1/2$.

\begin{figure}[h]
  \centerline{\includegraphics[width=\columnwidth]{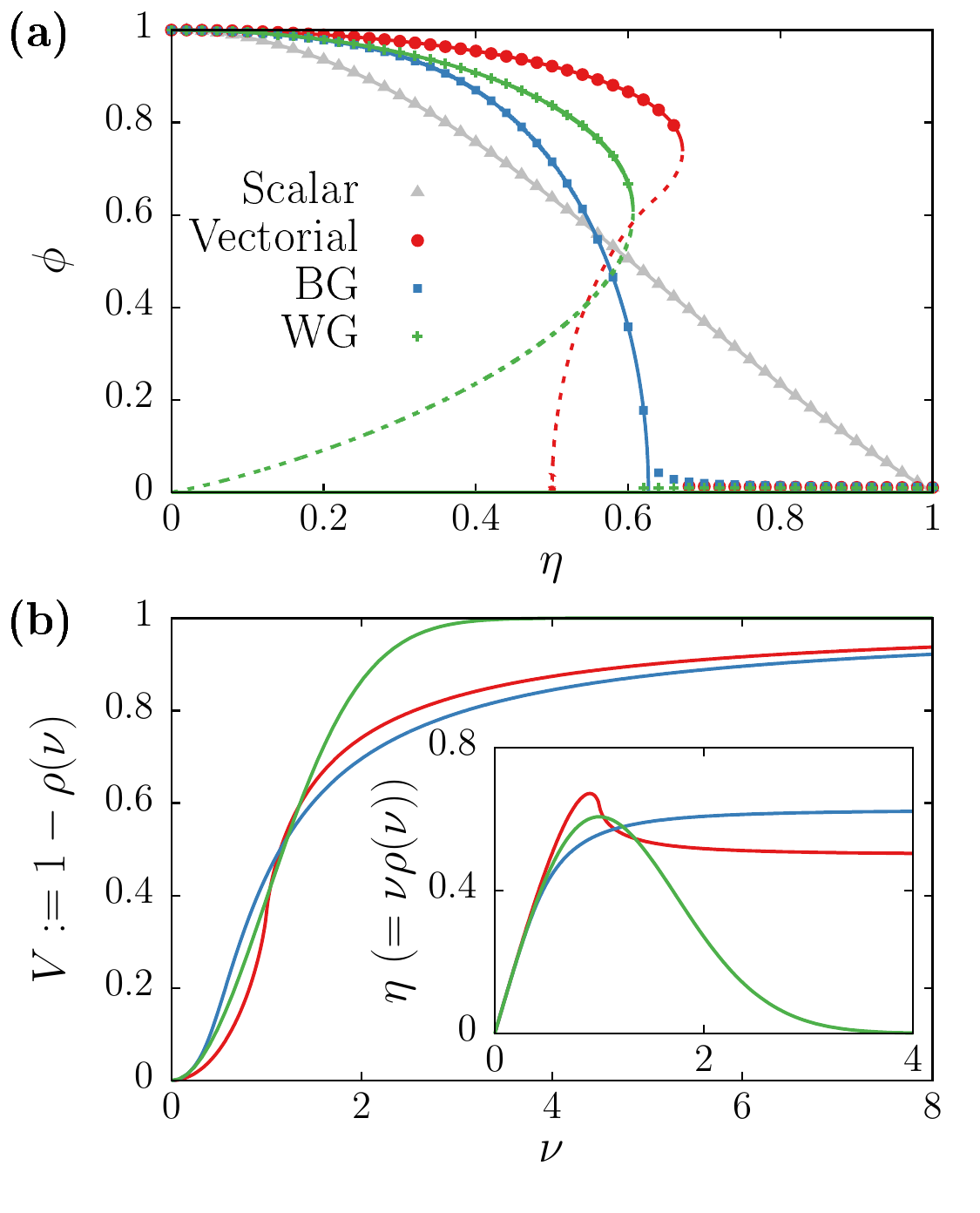}}
  \caption{(a) Mean-field transitions for the four different types of noise,
    namely original scalar noise (grey triangles), vectorial noise (red
    circles), bivariate Gaussian (blue squares), and wrapped Gaussian (green
    plusses).  Symbols correspond to simulations with $N=10^4$ particles
    starting from a fully ordered state.  Lines have been obtained from the
    analytical results.  A continuous line indicates a stable solution,
    whereas dashed lines correspond to unstable fixed points.  (b)
    Dependence of the circular variance, defined as $V=1-\rho$, on the
    parameter $\nu=\eta/\phi$ for the three multiplicative distributions
    (same colors as in (a)).  Inset shows the dependence of the control
  parameter on the distribution parameter $\nu$.  }
        \label{fig:meanfield}
\end{figure}

\subsection{Nature of the transition in multiplicative noise models}
Surprisingly, despite being qualitatively quite similar, the three
distributions with multiplicative noise exhibit an order-disorder transition
with three different scenarios.  The Vicsek model with vectorial noise has a
first order bifurcation with a small region of bistability; the wrapped
Gaussian noise model shows a first order bifurcation coexisting with a
disordered state for all values of $\eta$; and the bivariate Gaussian noise
model presents a continuous, second order phase transition.  Such
differences are not obvious from the definition of each distribution.  In
fact, in all three cases the amplitude (variance) of the distribution
increases with $\nu$, being more prominent for the Wrapped Gaussian noise
(see Fig~\ref{fig:meanfield}(b)).

In order to understand \emph{a priori} which choices of the angular noise
distribution lead to each of these different cases, one should notice that,
from the mean-field equation, $\rho(\nu)=\phi$.  Hence, at the fixed points
different from the disordered state $\phi=0$, the distribution parameter
$\nu=\eta/\phi$ can be written as $\nu=\eta/\rho(\nu)$.  Therefore, the
mean-field solutions corresponding to the ordered phase appear as the values
of the control parameter as a function of $\nu$, $\eta=\nu\rho(\nu)$ (see
Eq.~\eqref{eq:meanfield_numeric_multiplicative}).  The inset of
Fig.~\ref{fig:meanfield}(b) presents the functions $\nu\rho(\nu)$ for all
the cases studied here.  These functions show whether for a fixed $\eta$
there are one or two additional solutions to the disordered case.  The
existence of a local maximum indicates the presence  of two ordered
solutions in the vicinity of the maximum, and thus the presence of a
first-order transition, whereas a monotonic function characterizes a
second-order transition.

The two discontinuous first-order transitions presented here emerge as a
saddle-node bifurcation of a system at the thermodynamic limit.  Moreover,
the order parameter behaves as
\begin{equation}
	\phi-\phi_c \propto (\eta_c-\eta)^\beta
  \label{eq:criticality}
\end{equation}
where $\phi_c$ is the height of the jump at the discontinuity and $\beta$ is
a characteristic exponent.  According to this observation, we are therefore
in front of a \emph{hybrid phase
transition}~\cite{Dorogovtsev2006,Goltsev2006}, that is, a phase transition
that exhibits a jump in the order parameter, as in first order transitions,
accompanied by a critical singularity, as in second order transitions. For
the Vicsek model with vectorial noise we estimate the value of the critical
exponent numerically as $\beta\simeq0.5$ (see Table \ref{tab:betas}).  For
the wrapped Gaussian noise, recalling that $\phi_c=\eta_c=e^{-1/2}$, the
critical exponent can be obtained as 
\begin{equation}
	\beta = \lim_{\phi\to\phi_c}
  \frac{\log\left(\phi-\phi_c\right)}{\log\left(\eta_c-\phi\sqrt{-2\log\phi}\right)}=
  \frac{1}{2}.
\end{equation}

\begin{figure*}[t]
  \centerline{\includegraphics[width=0.9\textwidth]{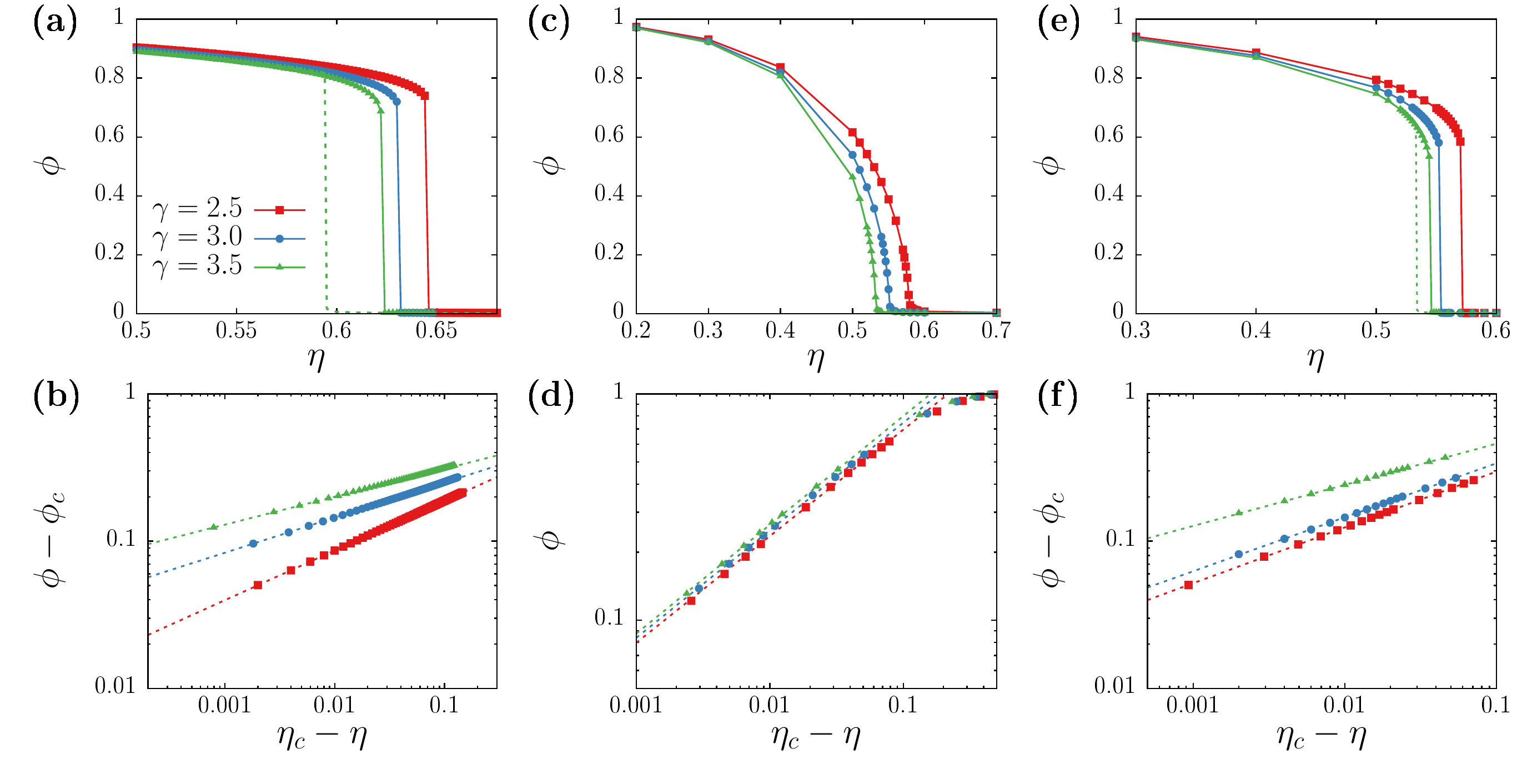}}
  \caption{Results of the generalized Vicsek models with fixed interactions
    given by power-law distributed  networks with $N=10^6$ nodes and degree
    distribution exponent $\gamma=2.5$ (red squares), 3 (blue circles), and
    3.5 (green triangles).  Top panels show the behavior of the order
    parameter $\phi$ for different values of $\eta$, whereas bottom panels
    show the corresponding scaling behavior close to the transition, with
    dashed lines indicating the results of a nonlinear fitting.  (a,b)
    Vectorial model.  (b,c) Bivariate Gaussian noise. (c,d) Wrapped Gaussian
    noise.  Dashed green lines in panels (a) and (c) display the results
    obtained by initializing the system at random for the case $\gamma=3.5$.
    All other results have been obtained starting from a fully ordered
    initial condition, with all SPPs pointing in the same direction.
  Simulations computed over a total of $2.5\cdot 10^5$ time steps, with a
thermalization of $5\cdot 10^4$ time steps.} 
        \label{fig:networks}
\end{figure*}

\section{Multiplicative noise models on networks}
\label{sec:networks}

In order to gain some understanding of the applications of mean-field theory
in our class of flocking dynamics models, we consider the case in which
interactions are mediated by a static complex network~\cite{Newman10}. The
relevance of this case is based on the recent claims that a network
structure, representing the patterns of social
interactions~\cite{croft2008exploring}, can be relevant in understanding the
flocking behavior of social
animals~\cite{Bode2011,Bode2011a,Miguel2018PRL,PhysRevE.100.042305,Ling2019},
which might have a stronger tendency to follow individuals with which they
have stronger social ties~\cite{Ling2019}.  Previous works have shown that
the scalar Vicsek model with networked interactions display a second-order
transition~\cite{Pimentel2008PRE,Miguel2018PRL}, whose properties depend on
the degree of heterogeneity of the network.  Here we present the results of
numerical simulations of generalized Vicsek models with multiplicative noise
on uncorrelated networks with a heterogeneous degree distribution given by a
power law $P(k) \sim k^{-\gamma}$, generated with the uncorrelated
configuration model (UCM)~\cite{Catanzaro2005PRE}. We consider networks of
size $N=10^6$ and different degree exponents $\gamma=2.5$, $3.0$, and $3.5$.
Our results are presented in Fig.~\ref{fig:networks}, where we display the
time averaged order parameter $\phi$ computed over a time span of $2\cdot
10^5$ time steps after a thermalization of $5\cdot 10^4$ time steps.
Results are computed starting from a completely ordered state, except for the
dashed green lines in Figures~\ref{fig:networks}(a,c), where the SPPs
initially point to randomly chosen directions.

In the case of the vectorial Vicsek model, see Fig.~\ref{fig:networks}(a),
we are clearly in presence of a first-order transition.  The critical value
$\eta_c$, at which the discontinuity of the order parameter takes place,
decreases with the degree exponent $\gamma$, but the overall discontinuous
character of the transition seems to be preserved. The hybrid nature of the
transition, predicted by the mean-field analysis, is still manifest on
networks, given by a critical singularity in the vicinity of the transition
in the ordered phase described by Eq.~\eqref{eq:criticality}.
Fig.~\ref{fig:networks}(b) shows this behavior, with the dashed lines being
the result of a non-linear fitting procedure.  A decrease of $\beta$ upon
increasing $\gamma$ can be appreciated.  Table~\ref{tab:betas} reports the
estimated values of $\beta$ for each case, as obtained from the non-linear
fitting.

For the case of the model with  bivariate noise, the transition displayed in
networks has a second-order nature, again in agreement with mean-field
theory, see Fig.~\ref{fig:networks}(c).  The critical exponent $\beta$
characterizing the transition appears to have a value roughly independent of
the critical exponent, and just slightly below the mean-field value $\beta =
1/2$, see Table~\ref{tab:betas} and Fig.~\ref{fig:networks}(d)).

The wrapped Gaussian case follows the same trend than the previous
instances: The behavior of the order parameter follows that of the
mean-field (see Fig.~\ref{fig:networks}(e)), including the hybrid behavior
close to the discontinuity.  In this case the numerically estimated values
of $\beta$ also change with $\gamma$ as in the vectorial model, see
Table~\ref{tab:betas} and Fig.~\ref{fig:networks}(f).

Finally, in order to illustrate the existence of multistability in the
vectorial and wrapped Gaussian noise, the green dashed lines in
Figs.~\ref{fig:networks}(a,c) show the order parameter obtained in
simulations starting from random initial conditions, in which the SPPs point
to randomly chosen directions. Only results for $\gamma=3.5$ are shown,
although the multistability of the vectorial and wrapped Gaussian noise is
present for all the tested networks.

\begin{table}[b]
  \caption{Estimated values of $\beta$ in mean-field and in power-law
    distributed networks with degree exponent $\gamma$. The mean-field
    values for the wrapped Gaussian and the bivariate Gaussian are
    analytical results.  The mean-field value for the vectorial has been
  obtained through a linear fitting of the log-log data.  The values for the
networks have been obtained through a non-linear fitting algorithm.}
	\label{tab:betas}

  \begin{ruledtabular}
  \begin{tabular}{ |c||c| c| c| c| }
	& Mean-field & $\gamma=2.5$ & $\gamma=3.0$ & $\gamma=3.5$  \\
 	\hline
  Vectorial & $0.489(3)$  & $0.34(2)$ & $0.24(2)$ & $0.19(5)$  \\
  Bivariate Gaussian  & 1/2 & $0.47(1)$ & $0.48(2)$ & $0.48(1)$ \\
  Wrapped Gaussian  & 1/2 &  $0.38(3)$ & $0.37(2)$ & $0.28(2)$  \\
\end{tabular}
\end{ruledtabular}
	\end{table}

\section{Multiplicative noise models on Euclidean space}
\label{sec:space}

\begin{figure}[t]
  \centerline{\includegraphics[width=0.9\columnwidth]{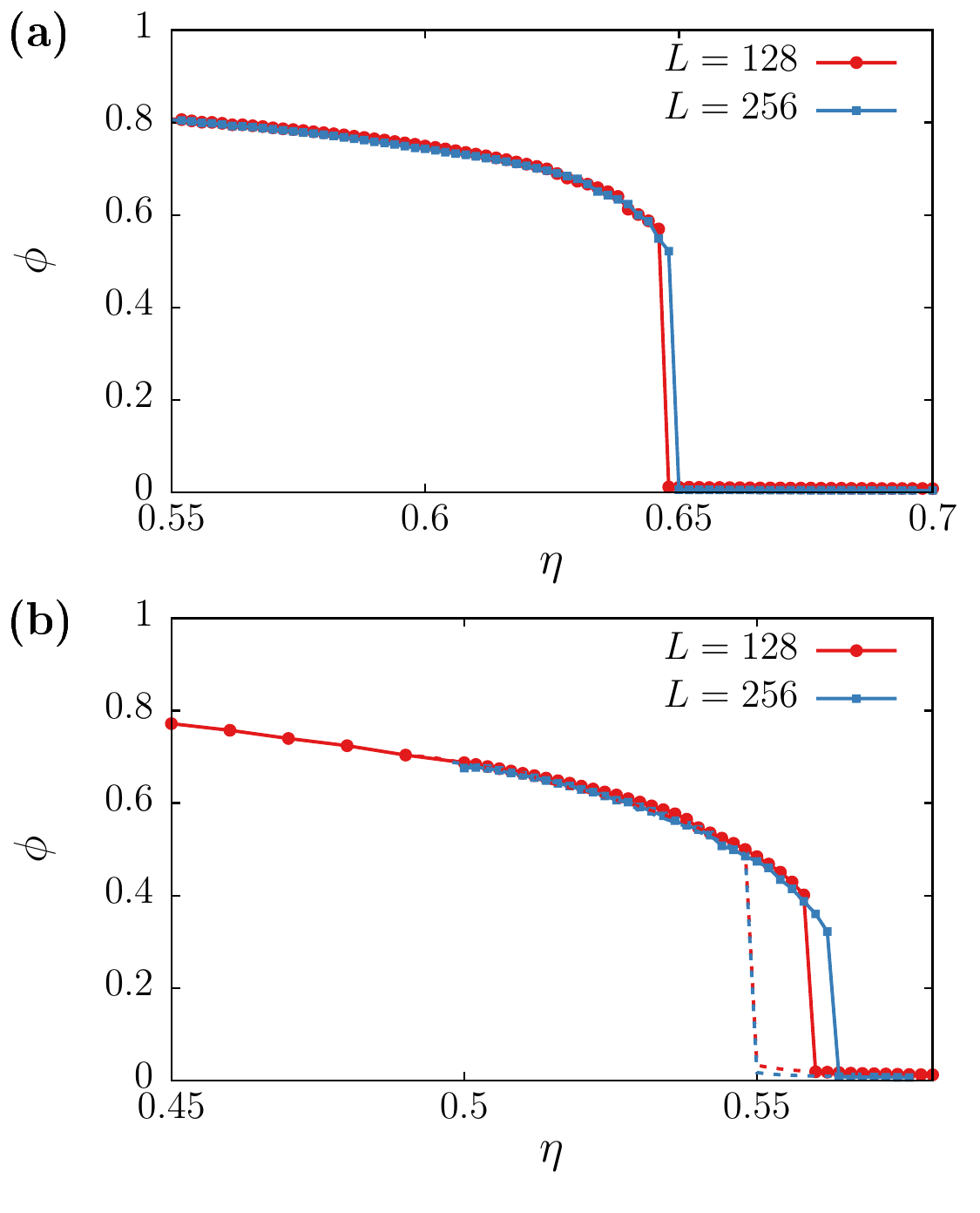}}
  \caption{Sample order parameter $\phi$ obtained from numerical simulations
    in 2D space with metric interactions.  (a) Results of the vectorial
    Vicsek model.  (b) Results of the Vicsek model with wrapped Gaussian
    noise.  Results correspond to systems with density $n=2$ and size
    $L=128$ (red circles) and $256$ (blue squares).  The length of the
    simulations is $2.5\cdot 10^5$ time steps including a thermalization of
    $5\cdot 10^4$ time steps.  Each simulation initial condition corresponds
    to the final configuration of the previous simulation. Continuous lines
    and points correspond to $\eta$ being increased, whereas dashed lines in
  panel (b) correspond to decreasing $\eta$.  }
  \label{fig:spatial}
\end{figure}

We finally study the effect of different multiplicative noise terms
%in the more realistic setting of a
in a two dimensional space in which the interactions
between SPPs are determined by the position of each unit.  The spatial
short-range interactions correspond to the formulation of the original
scalar Vicsek~\cite{Vicsek1995PRL}, in which interacting neighbors are
chosen following the metric rule in Eq.~\eqref{eq:metric}.  Nowadays it is
well known that both the original and vectorial formulations of the Vicsek
model with short-range spatial interactions display a first-order transition
with a coexistence region.  Whereas for the vectorial model this situation
is readily seen and in agreement with the mean-field results, the
first-order character of the original Vicsek model has been largely debated,
since the transition appears to be affected by strong finite-size
effects~\cite{Aldana2009IJMP,ChatePRL_comment2007PRL,Ginelli2016,Chate2008PRE,Aldana2009IJMP}.
Nonetheless, the evidence clearly indicates that, with periodic boundary
conditions, the transition from disorder to order remains discontinuous,
with a region of coexistence characterized by the formation of bands of high
density in the ordered
phase~\cite{Gregoire2004PRL,Ginelli2016,Chate2008PRE,chate2019dry}.  The
fact that the Vicsek model in the space presents a transition much different
from that seen in complex networks or the mean-field limit is due to the
strong effects caused by the interplay between density and coupling arising
from the short-range spatial interactions.  Changing the way in which such
short-range interactions are established can lead to qualitative different
behavior.  This is the case, for instance, of non-metric
interactions~\cite{Ginelli2010PRL}.  Therefore, it is \emph{a priori}
unclear which behavior we are going to observe in the different models of
multiplicative noise we consider here.

\begin{figure}[t]
  \centerline{\includegraphics[width=\columnwidth]{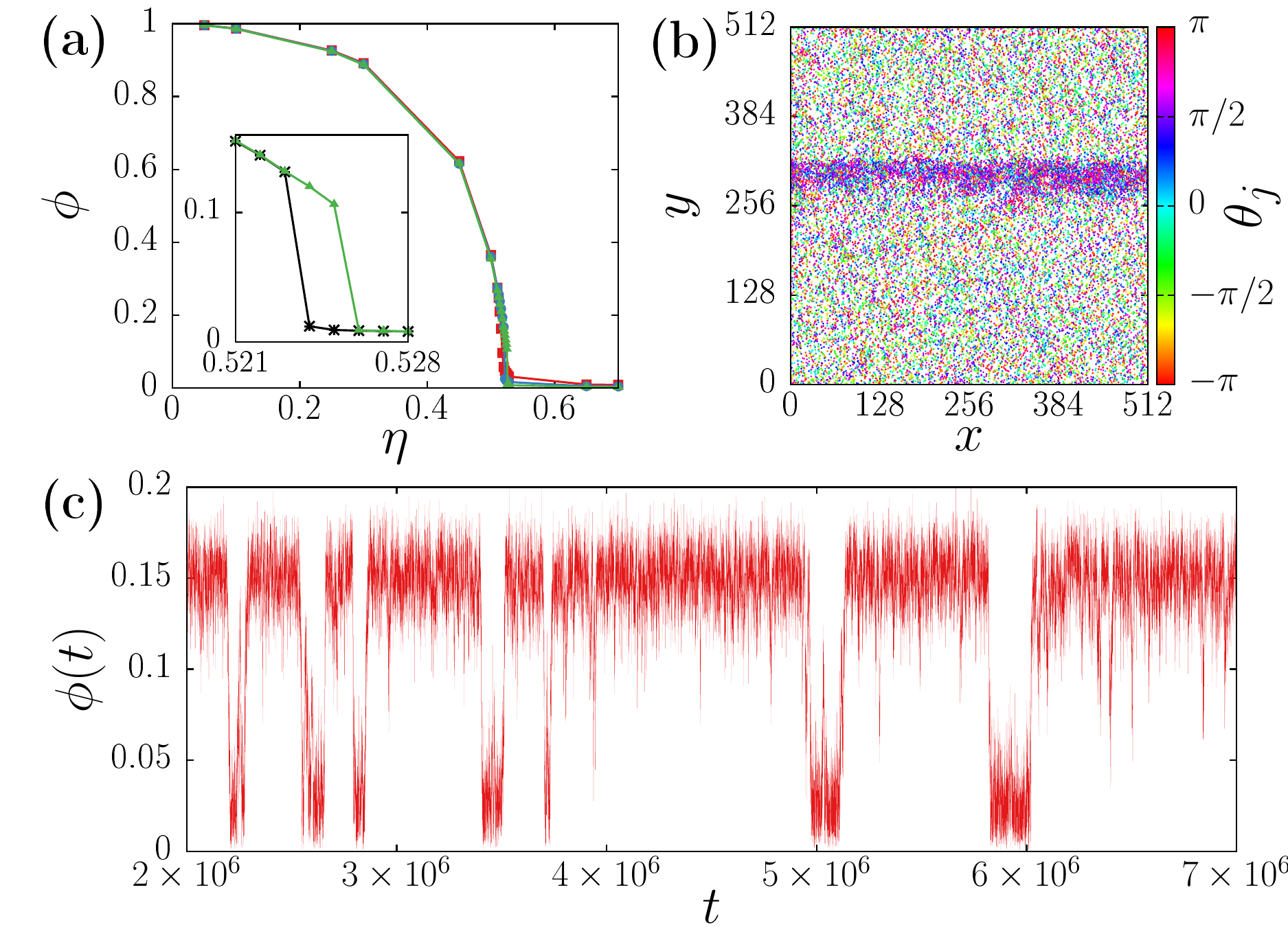}}
  \caption{(a) Order parameter $\phi$ obtained as a time average from
    numerical simulations of the model with spatial interactions and
    bivariate Gaussian noise. Red squares, blue circles, and green triangles
    correspond to simulations with a lattice of side length $L=128$, 256,
    and 512, with density $n=2$. The inset shows the hysteresis loop
    obtained by adiabatically increasing (green triangles) and decreasing
    (black stars) $\eta$.  Each simulation has been computed over a time
    window of $2.5\cdot 10^5$ time steps including a thermalization of $5
    \cdot 10^4$ time steps.  (b) Snapshot of the system at the ordered state
    for $L=512$, $n=2$, and $\eta=0.525$. The color of each particle
    indicates the orientation. Only $10\%$ of the particles are shown.  (c)
    Time series of the order parameter for $L=256$ and $\eta=0.521$ showing
  bistability between the ordered and disordered phases in a longer simulation.  }
  \label{fig:bivariate}
\end{figure}

In Fig.~\ref{fig:spatial} we display the results of numerical simulations
for vectorial and wrapped Gaussian models with metric interactions on a
square of size $L$ endowed with periodic boundary conditions.  In our
simulations the system size scales with $L$ so that the density of particles
is kept constant to the value $n=N/L^2=2$.  We used a radius of interaction
with neighboring particles of $r_0=1$ and the particles move at a speed
$v_0=0.5$.  The length of the simulations is $2.5\cdot 10^5$ time steps,
including a thermalization of $5\cdot 10^4$ time steps.  Simulations have
been performed adiabatically increasing $\eta$, except for dashed lines of
Fig.~\ref{fig:spatial}(b), which correspond to an adiabatic decrease of the
control parameter $\eta$.  
Both vectorial and wrapped Gaussian distribution
noise models display a clear discontinuous transition, with a region of
coexistence of ordered bands and disorder.  Beyond the coexistence region
the bands become unstable and the Toner-Tu ordered phase predominates.
Overall, the scenario is consistent with the usual bifurcation diagram of
the scalar Vicsek model~\cite{chate2019dry}.   In this case, the existence
of an hybrid scaling behavior at the transition point is not clear, as
simulations of the spatial systems display strong finite size effects.  

More interesting is the case of the bivariate Gaussian noise.
Fig.~\ref{fig:bivariate}(a) shows the results obtained for different system
sizes.  Although the transition appears to be continuous for small systems,
for large enough sizes ($L=256$ and $512$) it is possible to observe a
discontinuity on $\phi$, including an hysteresis loop (see inset of
Fig.~\ref{fig:bivariate}(a)).  Moreover, as reported in
Fig.~\ref{fig:bivariate}(b), snapshots of the simulations close to the
transition point show the existence of bands of ordered particles, a
signature of the classical bifurcation diagram of self-propelled polar
particles~\cite{chate2019dry}.  Finally, Fig.~\ref{fig:bivariate}(c) shows a
time series of the order parameter for $L=256$, where the bistability
between the ordered and disordered phases becomes evident.  Therefore,
despite finite-size effects, our results indicate that all spatial models
share a universal first-order transition towards the formation of ordered
bands, with a coexistence region.

\section{Discussion}
\label{sec:discussion}

Simple models of SPPs provide a framework to study the mechanisms underlying
the onset of self-organized collective motion.  In this paper we introduced
and studied a class of Vicsek-like models where the system dynamics depends
on two key elements: the definition of neighboring particles towards which
an individual tends to align, and a stochastic source of angular noise that
models the difficulties of a single individual to perfectly align with its
neighbors.  Previous studies considered different instances of interaction
rules\cite{Ginelli2010PRL,Bode2011,Miguel2018PRL}, but less work has been
devoted to the effect of different sources of angular noise on the onset of
collective motion. Two main stochastic angular noise terms have been used in
the literature: The uniform distribution introduced on the original Vicsek
model, and the vectorial noise proposed by Gr\'egoire and H.
Chat\'e~\cite{Gregoire2004PRL}, which, as we show, induces a multiplicative
effect.  Our derivation of the angular probability distribution for the
vectorial noise provides an explicit relation between the local polarization
of a SPP and the noise intensity, which appears to be mediated by the
parameter $\nu=\eta/a$.  Remarkably, the resulting probability density turns
out to be rather non-generic, with a piecewise distribution that is peaked
at the extremes for small noise values (or large local polarization), and
unimodal otherwise. 

We have extended the analysis of multiplicative noise in Vicsek-like models
by proposing two simpler and more generic angular probability distributions
for the stochastic source.  The case of bivariate Gaussian noise is
conceptually similar to that of the vectorial noise, where the angular
distribution needs to be computed \emph{a posteriori}. In this case, the
parameter $\nu=\eta/a$ also arises naturally from the definition of the
system.  On the other hand, the multiplicative character of the wrapped
Gaussian distribution is imposed with the definition of a standard deviation
that depends on the local polarization.

Despite being qualitatively similar, the differences between the three
multiplicative models studied here arises already at the mean-field level,
which is given in a generic form by solving a self-consistent equation
involving the mean-resultant length of the angular distribution.
Surprisingly, the character of the transition changes greatly with the
choice of the stochastic term, being a first-order hybrid transition for the
vectorial and wrapped Gaussian distribution, and second order for the
bivariate Gaussian model.  Simulations in complex networks show a scenario
consistent with that of the mean-field analysis, where different choices for
the noise leads to different behavior.

On the other hand, the emergence of collective motion in spatial models with
short-range interactions follows a mechanism of a different nature.  In all
the studied cases, numerical simulations with large enough number of
particles reveal a first-order transition including hysteresis and the
formation of high-density bands.  The scenario is reminiscent of that of the
original Vicsek model, where the phase transition for the spatial model does
not correspond to that seen in the mean-field or networked
models~\cite{chate2019dry,ChatePRL_comment2007PRL}.  Therefore we conclude
that the interplay between density and coupling that acts on spatial models
induces a rather generic phase diagram in which the peculiarities of the
noise distributions considered here play a minor role.

A single SPPs on a Vicsek-like model can be interpreted as an interacting
random walker in a two dimensional space.  Accordingly, the inclusion of
multiplicative noise in the model induces a correlation between the
direction of the next step and the state of the system. This idea brings the
multiplicative models studied here close to the model of active matter
through persistent random walkers proposed by Escaff \emph{et
al.}~\cite{Escaff2018} in one dimension.  In their model an individual
changes its direction only at certain time steps with a frequency that
depends on the local polarization.  In our case a change in direction
happens at all times, but it is much smaller if the local polarization is
high.  Thus we believe that a two dimensional model based on the idea of
persistent random walks should produce similar behavior than the one exposed
here, as well as could benefit from the analytical tools existing for that
case.

The study of minimal models provides solid grounds to study complex
phenomena and understand the effects of the different elements that compose
the model on the overall onset of collective motion.  On the other hand, it
is also important to work on realistic setups in order to retain new sources
for other types of phenomenology.  To this extend, possible extensions of
our work include considering continuous time models and allowing for steps
of different length (variable velocity)~\cite{Farrell2012PRL}, which can be
correlated with the local polarization and/or the random turning angle.
Finally, in line with the growing experimental work on the
field~\cite{19821081,Narayan2007,Sanchez2012,DeCamp2015,Li2018}, future work
should also study the validity of the different models on the basis of
empirical evidence.

\begin{acknowledgments}
  We acknowledge financial support from the Spanish MINECO, under Project
  No.  FIS2016-76830-C2-1-P, and Spanish MICINN, under Projects No.
  PID2019-106290GB-C21. 
\end{acknowledgments}

\section*{Data availability}

The network topologies used in the numerical simulations that support the findings of this study are available 
from the corresponding author upon request.

\bibliography{references}

\end{document}